\documentstyle[aps]{revtex}
\begin{document}
\draft
\title{Pure-radiation gravitational fields with a simple twist and a Killing vector}
\author{B.V.Ivanov}
\address{Institute for Nuclear Research and Nuclear Energy,\\
Tzarigradsko Shausse 72, Sofia 1784, Bulgaria}
\maketitle

\begin{abstract}
Pure-radiation solutions are found exploiting the analogy with the
Euler-Darboux equation for aligned colliding plane waves and the
Euler-Tricomi eqution in hydrodynamics of two-dimensional flow.They do not
depend on one of the spacelike coordinates and comprise the Hauser solution
as a special subcase.
\end{abstract}


\section{Introduction}

There exist many papers dealing with algebraically special, expanding and
twisting pure-radiation solutions of the Einstein equations. An extensive
bibliography up to 1980 exists in Ref. [1]. Further results on
pure-radiation fields can be found in Refs. [2-7]. The standard form of the
metric is [1] 
\begin{equation}
ds^2=\frac{2d\zeta d\bar \zeta }{\rho \bar \rho P^2}-2\Omega \left[
dr+Wd\zeta +\bar Wd\bar \zeta +H\Omega \right] ,  \label{1}
\end{equation}
\[
\Omega =du+Ld\zeta +\bar Ld\bar \zeta . 
\]

Here $r$ is the coordinate along the null congruence of geodesics, $u$ is
the retarded time, while $\zeta ,$ $\bar \zeta $ span a two-dimensional
surface. The metric components are determined by the $r$-independent real
functions $P,$ $m,$ $M$ and the complex function $L$

\begin{equation}
2i\Sigma =P^2\left( \bar \partial L-\partial \bar L\right) ,  \label{2}
\end{equation}
\begin{equation}
\rho =-\frac 1{r+i\Sigma },  \label{3}
\end{equation}

\begin{equation}
W=\rho ^{-1}L_u+i\partial \Sigma ,  \label{4}
\end{equation}
\begin{equation}
H=-r\left( \ln P\right) _u-(mr+M\Sigma )\rho \bar \rho +\frac K2,  \label{5}
\end{equation}
\begin{equation}
K=2P^2Re\left[ \partial \left( \bar \partial \ln P-\bar L_u\right) \right] ,
\label{6}
\end{equation}
where $\partial =\partial _\zeta -L\partial _u$ and $\Sigma $ is the twist.
The functions mentioned above satisfy the system of equations

\begin{equation}
\left( \partial -3L_u\right) \left( m+iM\right) =0,  \label{7}
\end{equation}
\begin{equation}
P^{-3}M=Im\partial \partial \bar \partial \bar \partial V,  \label{8}
\end{equation}
\begin{equation}
n^2=-2P^3\left[ P^{-3}\left( m+iM\right) \right] _u+2P^3\left( \partial
\partial \bar \partial \bar \partial V\right) _u-2P^2\left( \partial
\partial V\right) _u\left( \bar \partial \bar \partial V\right) _u,
\label{9}
\end{equation}
where $V_u=P$ , $n$ is the energy density of pure radiation and the Newton
constant is set to $1$. Equations (7)-(9) are in fact Eqs. (26.32) and
(26.33) from Ref. [1].

It has been noticed in different contexts that the condition $M=0$
simplifies the equations [4-6,8,9]. In a previous paper [10] we explored
this condition with the help of Stephani's method [2] for some algebraically
special, axisymmetric, expanding and twisting gravitational fields. They
depend on $u$ and $\sigma =\zeta \bar \zeta $ intrinsically i.e. the $u$%
-dependence cannot be taken away by applying the gauge transformations
(25.27) from Ref. [1]. In the present paper we find solutions for fields
which depend on $u$ and the real part of $\zeta ,$ $x=\frac 1{\sqrt{2}}%
\left( \zeta +\bar \zeta \right) $ and have the simplest possible twist.
Many of the well known algebraically special exact solutions possess this
kind of symmetry [1,3,4]. Among them are the only known seed vacuum
Robinson-Trautman solution of type III, the static C-metric and the only
known expanding vacuum solution of type $N$ with twist, namely the Hauser
solution, given respectively by Eqs. (24.15), (24.23) and (25.71) from Ref.
[1].

In Sec. II Eqs. (7)-(9) are reformulated in terms of an invariant potential
which leads to the $L_u=0$ gauge. In Sec. III solutions with separated
variables $u$ and $x$ are found. In Sec. IV the main Eq. (8) for simplest
twist is shown to be equivalent to a case of the Euler-Darboux equation,
somewhat different from the equation derived in Ref. [10]. On its turn it is
a complex version of the central equation in the theory of aligned colliding
plane waves (CPW). Three new solutions are found. In Sec. V a homogenous
hypergeometric solution is derived, exploiting the similarity between Eq.
(8) and the Euler-Tricomi equation. It contains as a special case the Hauser
solution of type $N$. Section VI contains some conclusions.

\section{Field equations in the $L_u=0$ gauge}

Following [2,10] we introduce the invariant complex potential $\phi $ which
solves Eq. (7): 
\begin{equation}
m+iM=\phi _u^3,  \label{10}
\end{equation}
\begin{equation}
L=\frac{\phi _x}{\sqrt{2}\phi _u}.  \label{11}
\end{equation}
When $M=0$ we can apply the transformation 
\begin{equation}
u^{\prime }=f\left( u,x\right) ,  \label{12}
\end{equation}
\begin{equation}
\left( m+iM\right) ^{\prime }=f_u^{-3}\left( m+iM\right)  \label{13}
\end{equation}
to make $m$ a positive or a negative constant $m_0$, so that 
\begin{equation}
\phi =m_0^{1/3}\left[ u+iq\left( x\right) \right] ,  \label{14}
\end{equation}
\begin{equation}
L=\frac i{\sqrt{2}}q_x  \label{15}
\end{equation}
with real $q$. Obviously $L_u=0$. This gauge differs from the usual gauge $%
P_u=0$ but is very suitable when the Newman-Unti-Tamburino (NUT) parameter $%
M $ vanishes. Equations (8) and (9) simplify 
\begin{equation}
\partial \partial \bar \partial \bar \partial V=\bar \partial \bar \partial
\partial \partial V,  \label{16}
\end{equation}
\begin{equation}
n^2=6m_0P^{-1}P_u+2P^3\partial \partial \bar \partial \bar \partial
P-2P^2\partial \partial P\bar \partial \bar \partial P  \label{17}
\end{equation}
with $\partial =\frac 1{\sqrt{2}}\left( \partial _x-iq_x\partial _u\right) $%
. The second equation is in fact an inequality. When $P_u\neq 0,$ $n^2$ can
be made positive by the choice of $m_0$ at least for some region of
spacetime [1,4,8]. The expressions for the metric components simplify too,
e.g., the gauge invariants $\Sigma $ and $K$ become 
\begin{equation}
\Sigma =\frac 12q_{xx}P^2,  \label{18}
\end{equation}
\begin{equation}
K=P^2\left( \bar \partial \partial +\partial \bar \partial \right) \ln P.
\label{19}
\end{equation}

When $m+iM=0$ (Petrov types III and $N$) Eq. (7) is an identity but still a
potential $\phi $ may be introduced with the property $\partial \phi =0$ and
the subclass of solutions satisfying Eq. (14) (with $m_0=1$ ) can be
studied. One should put $m_0=0$ in all other equations.

The main equation (16), which is of fourth order with respect to $V$,
becomes in both cases a linear second order equation for $P$. Equation (18)
shows that $q$ must be at least quadratic in $x$ for a non-trivial twist.
Let us choose the simplest possibility: $q=\frac{x^2}2,$ $L=\frac i{\sqrt{2}}%
x$. Then equations (16) and (17) read 
\begin{equation}
x^2P_{uu}+P_{xx}=0,  \label{20}
\end{equation}
\begin{equation}
2n^2=12m_0P^{-1}P_u-3P^3P_{uu}-4x^4P^2P_{uu}^2-P^2\left( P_u+2xP_{ux}\right)
^2.  \label{21}
\end{equation}
The last two terms in Eq. (21) are definitely negative, so the first term
must be necessarily positive for a type II solution and the second term must
be positive for a type III solution.

\section{Solution with separated variables}

Suppose that $P=F\left( x\right) G\left( u\right) $. Equation (20) splits
into two parts 
\begin{equation}
G_{uu}=cG,  \label{22}
\end{equation}
\begin{equation}
F_{xx}+cx^2F=0,  \label{23}
\end{equation}
where $c$ is an arbitrary constant. There are three types of solutions,
depending on the sign of $c$. If $c=0$ we have 
\begin{equation}
P=ux,  \label{24}
\end{equation}
\begin{equation}
n^2=\frac{6m_0}u-\frac 92u^2x^4.  \label{25}
\end{equation}
Type III solutions have negative energy while type II solutions have
negative energy density for $x\rightarrow \infty $. If $c>0,$ $P$ contains
Bessel functions, e.g., 
\begin{equation}
P=\sqrt{x}J_{1/4}\left( \frac{\sqrt{c}}2x^2\right) e^{-\sqrt{c}u},
\label{26}
\end{equation}
\begin{equation}
4n^2=-24\sqrt{c}m_0-2c\left( 3+4cx^4\right) F^4e^{-4\sqrt{c}u}-2cP^2\left(
F+2xF_x\right) ^2e^{-2\sqrt{c}u}.  \label{27}
\end{equation}
The first term can be made positive when $m_0<0,$ but the second has a
negative pole when $x\rightarrow \infty $ because $\left( xF\right) ^4\sim
x^2$. There are no solutions with everywhere positive $n^2$. The third case $%
c<0$ gives a generic solution with modified Bessel functions like 
\begin{equation}
P=\sqrt{x}K_{1/4}\left( \frac{\sqrt{-c}}2x^2\right) \sin \sqrt{-c}u,
\label{28}
\end{equation}

\begin{equation}
4n^2=24\sqrt{-c}m_0\cot \left( \sqrt{-c}u\right) -2c\left( 3+4cx^4\right)
F^4\sin ^4\sqrt{-c}u+2cP^2\left( F+2xF_x\right) ^2\cos ^2\sqrt{-c}u.
\label{29}
\end{equation}
The first term changes sign and has poles in $u$, thus type II solutions are
unphysical. Type III solutions have negative $n^2$ when e.g. $u=\pi /\sqrt{-c%
}$, although the second term doesn't have poles in $u$.

Equations (24) and (26) have been obtained by another method in a more
general form in Ref. [4] but the energy density has not been discussed in
details.

If we separate the variables like $P=F\left( x\right) +G\left( u\right) $
then 
\begin{equation}
G=-\frac{cu^2}2+c_1u+c_2,  \label{30}
\end{equation}
\begin{equation}
F=\frac{cx^4}{12}+c_3+c_4,  \label{31}
\end{equation}
where $c_i$ are constants. Numerous negative terms arise and the region of
positivity of $n^2$ is rather complicated due to the many arbitrary
constants. A more detailed discussion in a general setting of this type of
separation of variables may be found in Ref. [11].

\section{Reduction to the Euler-Darboux equation}

Equation (20) with $x^2$ replaced by $x^l$ where $l$ is an integer has been
studied in the past [12,13]. Green functions for different boundary problems
have been found. The basis of these results is the change of variables which
transforms Eq. (20) into the Euler-Darboux equation 
\begin{equation}
4P_{\varphi \omega }+\frac 1{\varphi +\omega }\left( P_\varphi +P_\omega
\right) =0,  \label{32}
\end{equation}
\begin{equation}
\varphi =u+\frac i2x^2,  \label{33}
\end{equation}
where $\omega =-\bar \varphi $.This equation is similar to Eq. (24) from
Ref. [10] and the main equation for aligned colliding plane waves [14] - one
needs only to replace the multiplier $4$ by $2$. This, however, is not a
trivial change and can't be achieved by scaling the variables or $P$.
Nevertheless, one may use the numerous techniques developed in the search
for CPW solutions and applied in Ref. [10] for axisymmetric fields. We shall
find analogues of these solutions. As pointed out in Ref. [10] the reality
of $P$ must be ensured because the variables $\varphi ,$ $\omega $ are
complex.

The simplest solution 
\begin{equation}
P=i^{-1/2}\left( \varphi +\omega \right) ^{1/2}=x  \label{34}
\end{equation}
is a time-independent vacuum solution of Kerr-Schild type.

A solution with separated variables $P=F\left( \varphi \right) G\left(
\omega \right) $ exists. Replacement in Eq. (32) yields 
\begin{equation}
P=A\left[ \left( \sigma +\varphi \right) \left( \sigma +\bar \varphi \right)
\right] ^{-1/4},  \label{35}
\end{equation}
where $A$ and $\sigma $ are constants. $A$ is ignorable and $\sigma $ can be
hidden in $u$ to obtain a real solution: 
\begin{equation}
P=B^{-1/4},  \label{36}
\end{equation}
\begin{equation}
B\equiv u^2+\frac 14x^4.  \label{37}
\end{equation}
This solution is analogous to the solution given by Eqs. (27) and (28) from
Ref.[10] and has a number of nice features. The energy density is 
\begin{equation}
n^2=-\frac{3m_0}{B^{3/2}}u+\frac 3{16B^3}\left( x^4-6u^2\right) -\frac 1{%
128B^5}\left[ 4x^4\left( 6u^2-x^4\right) ^2+u^2\left( 9x^4-4u^2\right)
^2\right] .  \label{38}
\end{equation}
In the following we suppose that the retarded time satisfies the condition $%
u>u_0>0$ for some constant $u_0$. Equation (38) is regular in $x$ unlike
many other solutions, plagued by singular pipes for $x=0$ or $x=\pm \infty $
[1]. When $m_0<0$ the first term dominates over the others if $\mid m_0\mid $
is big enough and consequently $n^2$ is positive. Unfortunately, type III
solutions are not with positive $n^2$ for any $x$ because the second term
changes sign. The gauge invariants (18) and (19) are regular in $x$ and
vanish when $u\rightarrow \infty $: 
\begin{equation}
\Sigma =\frac 12P^2,  \label{39}
\end{equation}
\begin{equation}
K=-\frac{x^2}{4B^{3/2}}.  \label{40}
\end{equation}
The same is true for the Weyl scalars [1,15] with leading terms given by 
\begin{equation}
\Psi _2=m_0\rho ^3,  \label{41}
\end{equation}
\begin{equation}
\Psi _3=-\rho ^2P^3\partial I+O\left( \rho ^3\right) ,  \label{42}
\end{equation}
\begin{equation}
\Psi _4=\rho P^2I_u+O\left( \rho ^2\right) ,  \label{43}
\end{equation}
\begin{equation}
I=P^{-1}\bar \partial \bar \partial P.  \label{44}
\end{equation}
An exception is $\Psi _2$ which approaches $-m_0/r^3$ as $u\rightarrow
\infty $.

Let us present next an analogue of the $\cosh ^{-1}$ solution found in Ref.
[10]. We substitute the ansatz $P=P\left( a\right) ,$ 
\begin{equation}
a=\frac{i\left( \varphi +\bar \varphi \right) }{\varphi -\bar \varphi }=%
\frac{2u}{x^2},  \label{45}
\end{equation}
into Eq. (32). The result is an elliptic integral of the first kind $F\left(
\psi ,\kappa \right) :$%
\begin{equation}
P=\sqrt{2}F\left( \psi ,\frac 1{\sqrt{2}}\right) ,  \label{46}
\end{equation}
\begin{equation}
\psi =\arccos \left( 1+a^2\right) ^{-1/4},  \label{47}
\end{equation}
\begin{equation}
P_a=\left( 1+a^2\right) ^{-3/4}.  \label{48}
\end{equation}
The function $P$ is bounded: $0\leq P\leq \sqrt{2}F\left( \frac \pi 2,\frac 1%
{\sqrt{2}}\right) $. The solution possesses regular characteristics like the
previous one: 
\begin{equation}
n^2=\frac{12m_0\mid x\mid }{P\left( 4B\right) ^{3/4}}+\frac{18u\mid x\mid P^3%
}{\left( 4B\right) ^{7/4}}-\frac{9x^2P^2}{4B^{3/2}},  \label{49}
\end{equation}
\begin{equation}
\Sigma =\frac 12P^2,  \label{50}
\end{equation}
\begin{equation}
K=-\frac 2{B^{1/2}}.  \label{51}
\end{equation}
All terms in Eq. (49) are regular. The first term is positive for $m_0>0$
and type II solutions with positive $n^2$ exist for big enough $m_0$. The
second term is positive too but does not always dominate over the third one.
The Weyl scalars are also regular and $\Psi _3,$ $\Psi _4$ vanish when $%
u\rightarrow \infty $.

Let us transform now the Euler-Darboux equation (32) into its canonical
form. Introducing the new variables $\tau =x^2,$ $\lambda =2u$ we obtain 
\begin{equation}
P_{\tau \tau }+\frac 1{2\tau }P_\tau +P_{\lambda \lambda }=0.  \label{52}
\end{equation}
This is an analogue of Eq. (25) from Ref. [10] and its solution is given by
the Bessel functions from Sec. III. We can go further, utilizing the
coordinates for the first Yurtsever solution [10,14,16]: 
\begin{equation}
\tau =\nu \sin \eta ,  \label{53}
\end{equation}
\begin{equation}
\lambda =\nu \cos \eta ,  \label{54}
\end{equation}
\begin{equation}
\nu ^2=4B,  \label{55}
\end{equation}
\begin{equation}
\cos \eta =\left( 1+\frac{x^4}{4u^2}\right) ^{-1/2}.  \label{56}
\end{equation}
Then Eq. (52) becomes 
\begin{equation}
P_{\nu \nu }+\frac 1{\nu ^2}P_{\eta \eta }+\frac 12\left( \frac 3\nu P_\nu +%
\frac 1{\nu ^2}\cot \eta ~P_\eta \right) =0.  \label{57}
\end{equation}
It has a separable solution of the kind $P=\nu ^lY\left( \eta \right) $. $Y,$
instead of being a Legendre function of the first or second kind, satisfies
the equation 
\begin{equation}
\left( 1-w^2\right) Y_{ww}-\frac 32wY_w+l\left( l+\frac 12\right) Y=0,
\label{58}
\end{equation}
where $w=\cos \eta $. Its solution is a hypergeometric function and 
\begin{equation}
P=\left( 4B\right) ^{l/2}F\left( \varepsilon ,\sigma ,-\frac 34,X\right) ,
\label{59}
\end{equation}
where $\varepsilon +\sigma =-\frac 52,$ $\varepsilon \sigma =l\left( l+\frac 
12\right) $ and 
\begin{equation}
X=\frac 12\left( 1+uB^{-1/2}\right) .  \label{60}
\end{equation}
$P$ is real because $0\leq X\leq 1$. The hypergeometric function is
reducible to a Legendre function: 
\begin{equation}
F\left( \varepsilon ,\sigma ,-\frac 34,X\right) =\Gamma \left( -\frac 34%
\right) \left( \frac{x^4}{4B}\right) ^{7/8}P_{-\frac{1+\varepsilon }2%
}^{7/4}\left( -uB^{-1/2}\right) .  \label{61}
\end{equation}
If $l=-\frac 12$, either $\varepsilon $ or $\sigma $ vanishes and formula
(59) degenerates to the rational function given by Eq. (36).

\section{Hydrodynamical analogy}

Equation (20) has certain similarities with the Euler-Tricomi equation 
\begin{equation}
x\Phi _{uu}-\Phi _{xx}=0.  \label{62}
\end{equation}
It appears in hydrodynamics in the study of a two-dimensional flow of
compressible fluid with velocity near the velocity of sound [17]. It is a
limiting case of the more complex Chapligin equation which has integrals
among the hypergeometric functions [17,18]. The Euler-Tricomi equation is
invariant under the transformations $u^2\rightarrow cu^2,$ $x^3\rightarrow
cx^3$ which leads to a homogeneous hypergeometric solution. It is discussed
at length in Ref.[17]. In our case Eq.(20) is invariant under $u\rightarrow
c^2u,$ $x\rightarrow cx$ and we can try a homogenous solution 
\begin{equation}
P=u^kF\left( z\right) ,  \label{63}
\end{equation}
\begin{equation}
z=-\frac{x^4}{4u^2},  \label{64}
\end{equation}
where $k$ is the degree of homogeneity. Then Eq. (20) becomes 
\begin{equation}
z\left( z-1\right) F_{zz}+\left[ \left( \frac 32-k\right) z-\frac 34\right]
F_z+\frac{k\left( k-1\right) }4F=0.  \label{65}
\end{equation}
Once again this is a hypergeometric equation and one of its fundamental
solutions leads to 
\begin{equation}
P=u^kF\left( -\frac k2,\frac{1-k}2,\frac 34,z\right) .  \label{66}
\end{equation}
In fact, the hypergeometric function in Eq.(66) degenerates to a Legendre
function for any $k$ 
\begin{equation}
P=2^{-1/4}\Gamma \left( \frac 34\right) u^k\left( -z\right) ^{1/8}\left(
1-z\right) ^{\frac k2-\frac 18}P_{-k-3/4}^{1/4}\left[ \left( 1-z\right)
^{-1/2}\right] .  \label{67}
\end{equation}
It becomes a rational function in some cases. Thus if $k=\frac 14$%
\begin{equation}
P=\left( \frac u2\right) ^{1/4}\left[ 1+\left( 1-z\right) ^{1/2}\right]
^{1/4},  \label{68}
\end{equation}
and if $k=-\frac 34$%
\begin{equation}
P=\left( \frac 2{u^3}\right) ^{1/4}\left( 1-z\right) ^{-1/2}\left[ 1+\left(
1-z\right) ^{1/2}\right] ^{1/4}.  \label{69}
\end{equation}

Exploiting only Eq. (64) one finds the following identities 
\begin{equation}
kP=uP_u+\frac 12xP_x,  \label{70}
\end{equation}
\begin{equation}
P_{uu}=P_1B_0^{-1},  \label{71}
\end{equation}
\begin{equation}
P_u+2xP_{xu}=\left( 4k-3\right) u^{-1}\left( kP-\frac 12xP_x\right)
-4uP_1B_0^{-1},  \label{72}
\end{equation}
\begin{equation}
P_1\equiv 4k\left( k-1\right) P-\left( 4k-3\right) xP_x.  \label{73}
\end{equation}
They hold for any solution of Eq. (65). Plugging Eqs. (70)-(73) into Eq.
(21) one can study the properties of the energy density. It simplifies
drastically when $k=\frac 34$:

\begin{equation}
P=u^{3/4}F\left( -\frac 38,\frac 18,\frac 34,z\right) ,  \label{74}
\end{equation}
\begin{equation}
\frac 16n^2=m_0P^{-1}P_u=\frac{m_0}{4u}\left[ 3-\frac{zF\left( \frac 58,%
\frac 98,\frac 74,z\right) }{2F\left( -\frac 38,\frac 18,\frac 34,z\right) }%
\right] .  \label{75}
\end{equation}
When $0\leq \mid z\mid \leq 1$ it can be checked with MAPLEV that the r.h.s.
of Eq. (75) is positive for $m_0>0$. In the region $1\leq \mid z\mid \leq
\infty $ an analytic continuation of the hypergeometric functions should be
performed. Thus 
\begin{equation}
F\left( -\frac 38,\frac 18,\frac 34,z\right) =A_1\left( -z\right)
^{3/8}F\left( -\frac 38,-\frac 18,\frac 12,\frac 1z\right) +A_2\left(
-z\right) ^{-1/8}F\left( \frac 18,\frac 38,\frac 32,\frac 1z\right) ,
\label{76}
\end{equation}

\[
A_1=\frac{\Gamma \left( \frac 34\right) \Gamma \left( \frac 12\right) }{%
\Gamma \left( \frac 18\right) \Gamma \left( \frac 98\right) }, 
\]
\[
A_2=\frac{\Gamma \left( \frac 34\right) \Gamma \left( -\frac 12\right) }{%
\Gamma \left( -\frac 38\right) \Gamma \left( \frac 58\right) }, 
\]
and similarly for the other hypergeometric function in Eq. (75). A check
with MAPLEV confirms again the positivity of $n^2$. The energy density is
regular in $u$ and $z$. When $m_0=0$ a vacuum solution is obtained. It has $%
\Psi _2=0$ and then the higher terms in Eq. (42) vanish [1]. From Eq. (44)
it follows that 
\begin{equation}
I=\frac 3{4\left( x^2-2iu\right) }  \label{77}
\end{equation}
and $\partial I=0,$ $I_u\neq 0$. Therefore $\Psi _3=0$, while $\Psi _4\neq 0$
and the solution is of type $N$. In fact, this is the Hauser solution [1,19]
in the $L_u=0$ gauge.

Let us show this in detail. We have used the transformation (13) to bring $%
\phi $ to the simple form (14) (with $m_0=1$ since the field is of type $N$)
and then have dropped the primes. Restoring them, Eqs. (12)-(14) show that

\begin{equation}
f=u^{\prime }=Cx^2u,  \label{78}
\end{equation}
where $C$ is an arbitrary constant which we fix to $C=1/2$. Under
transformation (12) $P$ and $L$ change as

\begin{equation}
P^{\prime }=f_u^{-1}P,  \label{79}
\end{equation}
\begin{equation}
L^{\prime }=f_uL-\frac 1{\sqrt{2}}f_x,  \label{80}
\end{equation}
(see Eq. (25.27) from Ref. [1]). $P^{\prime }$ is given by Eq. (74) while $%
L^{\prime }=ix/\sqrt{2}$. Then the original $P$ and $L$ are given by 
\begin{equation}
P=x^{7/2}F\left( u\right) ,  \label{81}
\end{equation}
\begin{equation}
L=\frac{\sqrt{2}}x\left( u+i\right) ,  \label{82}
\end{equation}
\begin{equation}
F\left( u\right) =\frac 12A_1F\left( -\frac 38,-\frac 18,\frac 12%
,-u^2\right) +\frac 14A_2uF\left( \frac 18,\frac 38,\frac 32,-u^2\right) .
\label{83}
\end{equation}

We have used formula (76) to derive the expression for $F\left( u\right) $.
Equations (81)-(83) give exactly the Hauser solution [19] as written out in
Refs. [1] and [20]. Equation (83) is a specific linear combination of the
even and odd solutions given in Ref. [20] which fixes the one-parameter
freedom of the Hauser solution. This ends the proof of our assertion.

\section{Conclusion}

We have shown that when the NUT parameter $M$ vanishes and the gauge $L_u=0$
is used, the main equation (20) for expanding pure radiation fields with a
simple twist and a special symmetry becomes a tractable second order linear
equation for $P$. It is reducible to a case of the Euler-Darboux equation,
somewhat different from the central equation in the theory of aligned
colliding plane waves. We have found regular solutions, separating the
variables in different coordinate systems. In some cases the regions of
positivity of the energy density were investigated. Another analogy with the
Euler-Tricomi equation, appearing in the hydrodynamics of two-dimensional
fluid flow, has been exploited to find homogenous solutions. Interestingly
enough, the Hauser vacuum solution of type $N$ is an exceptional member of
the family of type II solutions with degree of homogeneity $\frac 34$.

\section*{Acknowledgments}

This work was supported by the Bulgarian National Fund for Scientific
Research under Contract No. F-632.


\begin{references}
\bibitem{one}  D. Kramer, H. Stephani, M. MacCallum and E. Herlt, {\it Exact
Solutions of Einstein's Field Equations }(Cambridge University Press,
Cambridge, England, 1980).

\bibitem{two}  H. Stephani, Gen. Relativ. Gravit. {\bf 15}, 173 (1983).

\bibitem{three}  J. Lewandowski, P. Nurowski, and J. Tafel, Class. Quantum
Grav. {\bf 8}, 493 (1991).

\bibitem{four}  J. Tafel, P. Nurowski, and J. Lewandowski, Class. Quantum
Grav. {\bf 8}, L83 (1991).

\bibitem{five}  A. Grundland and J. Tafel, Class. Quantum Grav. {\bf 10},
2337 (1993).

\bibitem{six}  D. Kramer and U. H\"ahner, Class. Quantum Grav. {\bf 12},
2287 (1995).

\bibitem{seven}  U. von der G\"onna and D. Kramer, Class. Quantum Grav. {\bf %
15}, 2017 (1998).

\bibitem{eight}  H. Stephani, J. Phys. A {\bf 12}, 1045 (1979).

\bibitem{nine}  E. Herlt and H. Stephani, Class. Quantum Grav. {\bf 1}, 95
(1984).

\bibitem{ten}  B. V. Ivanov, Phys. Rev. D {\bf 60}, 104005 (1999).

\bibitem{eleven}  B. V. Ivanov, J. Math. Phys. {\bf 40}, (1999), to be
published.

\bibitem{twelve}  E. Holmgren, Ark. Mat. Astr. Fys. B{\it \ }{\bf 19, }1
(1926).

\bibitem{thirteen}  S. Gellerstedt, Ark. Mat. Astr. Fys. A{\it \ }{\bf 25},%
{\bf \ }1 (1936).

\bibitem{fourteen}  J. B. Griffiths, {\it Colliding Plane Waves in General
Relativity }(Clarendon,Oxford, 1991).

\bibitem{fifteen}  D. W. Trim and J. Wainwright, J. Math. Phys.{\it \ }{\bf %
15},{\bf \ }535 (1974).

\bibitem{sixteen}  U. Yurtsever, Phys. Rev. D{\it \ }{\bf 37},{\bf \ }2790
(1988).

\bibitem{seventeen}  L. D. Landau and E. M. Lifshitz, {\it Hydrodynamics }%
(Nauka, Moscow, 1988).

\bibitem{eighteen}  R. Mises, {\it Mathematical Theory of Compressible Fluid
Flow }(Academic Press, New York, 1958).

\bibitem{nineteen}  I. Hauser, Phys. Rev. Lett.{\it \ }{\bf 33},{\bf \ }1098
(1974).

\bibitem{twenty}  D. E. Novoseller, Phys. Rev. D {\bf 11}, 2330 (1975).
\end{references}
\end{document}